\documentclass[authoryear,preprint,12pt]{elsarticle}

\usepackage{amssymb}
\usepackage{amsmath}
\usepackage{tabularx}
\usepackage{longtable}
\DeclareMathOperator{\argmax}{arg\,max}
\usepackage{tikz}
\usepackage{hyperref}
\usetikzlibrary{calc}
\usepackage[font=footnotesize]{caption}

\begin{document}

\hypersetup{pageanchor=false}
\begin{titlepage}
    \begin{center}
        \vspace*{0.1cm}
        {\LARGE From transient shocks to unexpected outcomes: disruptive drivers in scenario pathways\par}
        \vspace{1.0cm}
Andrew G. Ross $^{1,*}$ \\
        \vspace*{1.25\baselineskip}
         $^1$ Forschungszentrum J\"ulich GmbH, Institute of Climate and Energy Systems, J\"ulich Systems Analysis, Wilhelm-Johnen-Strasse, 52425 J\"ulich, Germany \\
        \vspace*{0.5\baselineskip}

        \vspace*{1\baselineskip}
\vspace*{2\baselineskip}
$^*$ Corresponding author: \url{a.ross@fz-juelich.de} \\
\vspace*{2\baselineskip}
\today

\vspace*{1.25\baselineskip}

\textcolor{red}{\textbf{NOT PEER-REVIEWED VERSION}}

    \end{center}

\end{titlepage}

\begin{titlepage}
    \begin{center}
        \vspace*{0.1cm}
        {\LARGE From transient shocks to unexpected outcomes: disruptive drivers in scenario pathways\par}
       % Transient events, disruptive drivers, and unexpected outcomes in scenario pathway modelling\par}
        \vspace{0.5cm}
    \end{center}

    \vspace{0.5cm}

    {\noindent Scenario pathways (e.g. for the energy transition) often use a single trajectory or a band. That is not sufficient when one needs to understand why outcomes differ and under what stress or uncertainty they arise. Doing so requires tracking disequilibrium along pathways, comparing runs across ``worlds'' or storylines, and surfacing outcomes that are unlikely under a central view but plausible when how factors interact is uncertain. Cross-Impact Balance (CIB) is a well-established method for generating pathways. This paper extends CIB to formalise and implement these dimensions in pathway runs, and defines four run types that respectively emphasise one-off shocks, extremes under alternative regimes, influence-structure uncertainty that widens over time, and exogenous shocks as a baseline for comparison. The approach is applied to a socio-technical decarbonisation pathway for illustration. Together, the extensions support stress-testing, comparison across storyline or regime assumptions, and exploration of rare or surprising futures, and help analysts distinguish results that are stable across those assumptions from those that depend on structural uncertainty about the influence table.} \\

    \vspace{0.5em}

    \vspace{0.5em}


    \vfill

\end{titlepage}
\hypersetup{pageanchor=true}

\newpage

%%%%%%%%%%%%%%%%%%%%%%%%%%%%%%%%%%%%%%%%%%%%%%%%%%%%%%%%%%%%%%%%%%%%%%%%%%%%%%%%%%%%%%%%%%%%%%%%%%%%%%%%%%%%%%%%%%%%%%%%%%%%%%%%%%%%%%%%%%%%%%%%%%
\section{Introduction}
\label{sec1}

\noindent Scenario and pathway analysis is widely used to explore how the future might unfold under different assumptions. Frameworks such as the Shared Socioeconomic Pathways provide qualitative narratives that describe world futures and are combined with quantitative pathways for integrated assessment \citep{RIAHI2017153,ONeill2017}. Similar approaches are used in energy and climate context scenarios, where storylines are built from consistent factor combinations and then linked to various assessment models \citep{Rounsevell2010,Guivarch2017}.\\

\noindent Nevertheless, several dimensions remain underexplored in practice. These are classified following \citet{McCollum2020,Weitzman2009}. First, the quantification of short-lived departures from equilibrium along pathways is methodologically underdeveloped. That is, policy and strategy would benefit from knowing how far a pathway is from a stable, self-consistent state at each step. Second, the exploration of pathways under alternative structural assumptions such as disruptive drivers or extremes is often limited. Comparing outcomes under distinct ``worlds'' or storylines is not routinely separated from other sources of variation. Third, the exposure of rare or overlooked outcomes when the influence structure is uncertain receives insufficient attention, including tail events and what has been termed grey or black swans \citep{Taleb2007}.\\

\noindent When scenario work informs policy or strategy, a single central trajectory or a single band of outcomes is often insufficient \citep{Pindyck2013,ROSEN2025103566}. Decision-makers need to understand not only how much outcomes can vary but why they differ and under what kind of stress or uncertainty they arise. A min-max band added to a central trajectory shows the range of outcomes when inputs are varied, but it does not separate the effect of a one-off shock from the effect of a different storyline or from the effect of uncertainty about how the system works. The band remains an uninterpretable range. To the best of the author's knowledge, existing scenario pathway and storyline work (including overviews of quantitative scenario techniques \citep{Guivarch2017}) does not provide a single framework that separates these influences so that the kind of stress or storyline or uncertainty behind variation can be identified and communicated \citep{lempert2003shaping,McCollum2020}. The pathway framework set out in this paper supports such attribution when scenarios are built as distinct runs: shock, regime, uncertain influence structure, and exogenous disturbances can be turned on or combined separately. Where an illustration bundles several of these, interpretation follows that joint setup.\\

\noindent Cross-Impact Balance (CIB) provides a framework in which these gaps can be addressed. At its core is a cross-impact matrix encoding influences between descriptor states (descriptors are factors, each with a set of discrete ordinal states) \citep{WEIMERJEHLE2006334}. With multi-period pathways it is commonly used for the generation of scenario pathways \citep{schweizer2012improving,WeimerJehle2020,Schweizer2020}. This paper extends CIB along three dimensions: disequilibrium along pathways (short-lived departures from a stable state), pathways under distinct regimes or ``worlds'' (extremes, where a regime is a named influence structure under which pathways are run), and unexpected outcomes when the influence table is uncertain.\\

\noindent In this paper, a systematic approach is set out that extends CIB in this way. Four scenario types are developed and illustrated, three of which build on the new extensions and one on existing stochastic exogenous shocks retained for comparison. Each type highlights a different cause of variation (transient shock, a joint regime-and-shock specification, system uncertainty, or random disturbances alone), so that pathways are explored in four distinct ways and both the extent and the kind of causes of variation in outcomes are visible. The distinction is qualitative. Each scenario type is a distinct combination of mechanisms so numerical differences between types are not interpretable as the quantitative effect of a single factor. The quantification of disequilibrium along pathways is formalised here and is not provided by other CIB or scenario pathway methods in this form. Formal disequilibrium metrics (consistency margin, burden, partial convergence) and pathway-level sampling from an uncertain CIM are not offered in existing CIB pathway or scenario overviews \citep{Guivarch2017}.\\

\noindent The four scenario types are: runs under a one-off stronger disturbance in a single period (stress-testing and resilience), runs under a joint regime-and-shock specification (comparison across such joint specifications), runs under uncertain influence structure (rare or surprising endpoints), and runs with exogenous stochastic disturbances only (baseline of pure noise). Including this fourth type as a distinct run provides a baseline against which the added effect of shock, the joint regime-and-shock specification, and structural uncertainty can be compared. Together they support stress-testing, comparison across joint regime specifications, and the identification of rare or surprising futures.\\

\noindent The three extensions to CIB (disequilibrium, extremes, and unexpected outcomes) are the main contribution of this paper and the first to combine these three dimensions within a single CIB pathway framework. The approach is general and applicable to any CIB application. For illustration it is applied here to a socio-technical pathway based on the data of \citet{ROSSROSS}, which uses AI-simulated expert panels for its generation. Each scenario type can involve tail-like or rare outcomes in a different sense (realisation tails, tails under the joint extremes specification, or structural-uncertainty tails). The four types together allow one to assess whether results are robust across joint regime specifications, depend on the chosen joint extremes specification, or only appear when structural uncertainty is included \citep{lempert2003shaping,Taleb2007}.\footnote{Thematically, realisation tails can also surface rare or disruptive pathway outcomes in the spirit of grey swans (plausible but often neglected). In this paper the terms ``grey swan'' and ``black swan'' are reserved for the third scenario type (structural-uncertainty tails) so that the cause of rarity remains clearly attributed.}\\

\noindent This paper is organised as follows. Section~\ref{sec:conceptual} provides further non-technical detail on the scenario literature, the evidence for the limitation of single-band presentation, and the four scenario types and their interpretation. Section~\ref{sec2} outlines the CIB method and the novel extensions in formal terms. Section~\ref{sec2:simulation} describes the simulation scenario and points to the appendices which specify the simulation scenario and notation. Section~\ref{Simulation_results} reports simulation outcomes for the four scenario types. Section~\ref{Discussion} discusses and concludes.

%%%%%%%%%%%%%%%%%%%%%%%%%%%%%%%%%%%%%%%%%%%%%%%%%%%%%%%%%%%%%%%%%%%%%%%%%%%%%%%%%%%%%%%%%%%%%%%%%%%%%%%%%%%%%%%%%%%%%%%%%%%%%%%%%%%%%%%%%%%%%%%%%%
\section{Conceptual background}
\label{sec:conceptual}

\noindent The following develops the conceptual basis outlined in the introduction. It adds the scenario and pathway literature, the limitation of single-band presentation, and the four scenario types and their interpretation. Section~\ref{sec2} then provides the technical detail.\\

\noindent The scenario pathways and storylines literature stresses trajectories over time and narrative assumptions about alternative ``worlds'' to explore how the future might unfold. Reviews of scenario techniques for energy and environmental research emphasise the need to broaden the capacity to deal with complexity and uncertainty and to combine qualitative and quantitative elements \citep{Guivarch2017,KWAKKEL2013419,BAI2016351}. That literature supports comparison across narrative assumptions and stress-testing of strategies under different futures \citep{lempert2003shaping}.\\

\noindent The gaps outlined in the introduction are reflected in this literature. Calls to explore extremes more systematically note that the future is shaped as much by extremes as by central trends \citep{McCollum2020,Weitzman2009}. Work on resilient climate mitigation, for example, argues for expanded futures analysis rather than relying on conventional modelling alone \citep{Gambhir2023}. Broader exploration of the scenario space is supported by evidence that many scenarios are needed for robust energy system insights and that single or narrow scenario sets can miss important outcomes \citep{Frey2025}. Representativeness of scenario sets is argued to require attention to ``unknown unknowns'' and to structural uncertainty in how factors interact \citep{ERIKSSON2022102939,Walker01032003}. The extensions developed in this paper contribute to that representativeness.\\

\noindent The limitation of presenting outcomes as a single band has been widely discussed \citep{McCollum2020}. A min-max band does not separate the effect of a one-off shock from that of a different storyline or from uncertainty about how the system works. In risk assessment, nonprobabilistic cross-impact approaches have been criticised for not guaranteeing conservative risk estimates and for consistency criteria that can admit very few or no scenarios \citep{Salo2022}. Such simplified approaches are widespread in scenario pathway practice. \citet{McCollum2020} note that they are (too) common in energy scenario modelling.\\

\noindent CIB is commonly used for the generation of multi-period scenario pathways \citep{schweizer2012improving,KEMPBENEDICT201955,Schweizer2020}. Pathways evolve from a cross-impact matrix over a sequence of periods so that each period yields a scenario. Linked together, these scenarios form a pathway. The approach is used across many domains, including socio-technical scenarios for net-zero and other transformation pathways, and supports scenarios that combine qualitative storylines with quantitative models in plausible societal contexts for model assessment \citep{WeimerJehle2020, Guivarch2017}. Existing CIB pathway work has focused on consistency and multi-scale linking \citep{Guivarch2017,KEMPBENEDICT201955}.\footnote{Other cross-impact approaches derive scenario probabilities and integrate with risk or decision analysis \citep{Salo2022,Roponen2024}, but typically do not provide pathway-level dynamics, disequilibrium metrics, or the separation of shock, storyline, and structural uncertainty in a single run framework \citep{Guivarch2017}. That overview and the CIB developments it discusses (e.g.\ linked CIB for multi-scale consistency \citep{KEMPBENEDICT201955}) address different dimensions rather than pathway-level disequilibrium or uncertain-CIM sampling.} The extensions below add dimensions that this literature does not yet address.\\

\noindent Within this CIB pathway framework, four scenario types developed in this paper serve distinct analytical purposes. The following discusses each in turn.\\

\noindent The first of the four scenario types addresses what happens when the system is subjected to a one-off stronger disturbance in a single period. Its purpose is stress-testing, namely to make visible the response to a short-lived disruption and to quantify how far the pathway is from a stable, self-consistent CIB state at each step. The insights are the resilience or vulnerability of pathways to an amplified event and the extent of transient disequilibrium along the path. Outcomes can include rare realisations when shock sequences are extreme (realisation tails). Decision-makers can use this to identify where buffering or contingency planning is most needed.\\

\noindent The second type addresses what happens when the analyst assumes a different underlying narrative or structural assumption (for example a baseline, moderate, or stress ``world'') and holds it fixed for the whole pathway. Each storyline is represented by a distinct influence structure (regime).\footnote{Defining extremes as regime-based storylines (alternative CIMs) rather than as events or as diversity bounds (as commonly done) under a single CIM supports comparison of robustness versus storyline-dependence and makes stress-testing under adverse influence structures a distinct, interpretable scenario type in principle. As \citet{ROSSROSS} highlight, in much existing CIB work that refers to extremes, the term denotes either extreme events (e.g.\ droughts, floods, or other rare high-impact occurrences) or the maximally diverse scenarios that bound the scenario space under one matrix. Regime-based extremes as a scenario type are not modelled in that way. The approach taken here is therefore clearly distinct from that literature. In the general formalism, regime (fixed CIM), structural shocks, and uncertain-CIM sampling are separate optional layers. In the empirical illustration, extremes pathway ensembles combine regime-specific matrices with regime-specific structural shock scaling.} The purpose is to compare outcomes across storylines and to assess sensitivity to the choice of narrative. The insights are which outcomes are robust across alternative regimes or narratives and which depend on the assumed world, which is useful in contested policy contexts.\footnote{Robustness means similar outcome distributions under different assumed regimes or narratives. Material dependence means those distributions differ across assumptions \citep{Guivarch2022,Lempert2006}. Structural shock intensity and other disturbance settings may be specified per regime. Where they co-vary with the regime matrix, pathway contrasts reflect that joint specification. Attribution to narrative or influence structure alone requires that such settings be held fixed across regimes or varied in a separate factorial design.} This type explores narrative or stress tails (outcomes under deliberately adverse or ``tail'' storylines).\\

\noindent The third type addresses what happens when the influence table is not treated as known but is instead sampled from an uncertain specification, with uncertainty optionally widening over time and disturbances applied. The purpose is to expose pathway endpoints that are rare under a single central matrix but plausible under structural uncertainty, including tail-risk and grey-swan outcomes \citep{Walker01032003,Taleb2007}. The insight is the share and identity of such rare but plausible states (endpoints a central matrix may underweight) when the influence structure is uncertain (structural-uncertainty tails). The aim is to broaden the set of futures considered and to avoid overconfidence in a single view. \\

\noindent The fourth type uses a single nominal backbone CIM and applies only exogenous structural and dynamic stochastic disturbances, without uncertain-CIM sampling or regime change (the realised matrix still varies period-to-period when structural shocks are drawn). Its purpose is to isolate the exogenous shock channels from matrix-sampling uncertainty and storyline regimes, and thereby to provide a reference workflow for qualitative contrast with the other three types. It can yield realisation tails when shock sequences are extreme. This scenario type is outlineed in \cite{ROSSROSS} and thereby not developed further in this paper. The formal treatment of the four types and their implementation is given in the following section.\\

%%%%%%%%%%%%%%%%%%%%%%%%%%%%%%%%%%%%%%%%%%%%%%%%%%%%%%%%%%%%%%%%%%%%%%%%%%%%%%%%%%%%%%%%%%%%%%%%%%%%%%%%%%%%%%%%%%%%%%%%%%%%%%%%%%%%%%%%%%%%%%%%%%
\section{Methods}
\label{sec2}

\noindent The following subsections outline the method in sequence. The standard CIB framework and multi-period pathways, then three extensions (disequilibrium, extremes, and unexpected outcomes), each defined and described in terms of the implementation within PyCIB (an open-source CIB analysis package) \citep{PyCIB}. The purpose and interpretation of each scenario type, and the distinction between realisation, narrative, and structural-uncertainty tails, are as set out in Section~\ref{sec:conceptual}. The following subsections give the formal definitions and implementation.\\

\subsection{Standard CIB and multi-period pathways}
\label{sec2:cib}

\noindent At the core of the CIB method is a cross-impact matrix (CIM), $C$, the table of pairwise influence scores between descriptor states. Descriptors are factors, each with a set of discrete ordinal states. Following \cite{WEIMERJEHLE2006334}, a scenario $z$ is one state chosen per descriptor (i.e.\ one state per factor), and $z_j$ denotes the state of descriptor $j$ in scenario $z$. For such a scenario, the impact score of target descriptor $j$ in state $l$ is

\begin{equation}
\label{eq:impact}
\theta_{j,l}(z) = \sum_{i \neq j} C_{i \to j}(z_i, l),
\end{equation}

where $i$ runs over descriptors, $z_i$ is the state of source descriptor $i$ in scenario $z$, $l$ denotes a possible state of descriptor $j$, and $C_{i \to j}(z_i, l)$ is the impact of source descriptor $i$ in state $z_i$ on target descriptor $j$ in state $l$.\\ 

\noindent In regime-aware runs the matrix used in this formula is the active matrix $M_{r(t)}$ for period $t$ (Section~\ref{sec2:extremes}); otherwise it is the single CIM $C$. The impact balance for descriptor $j$ is the vector of these scores over all its states. A scenario is consistent if, for every descriptor $j$, the chosen state $z_j$ attains the maximum (ties allowed). The implementation treats a descriptor as inconsistent only if some other state has strictly greater score.\footnote{In this application, ``consistency'' means the CIB rule (Eq.~\eqref{eq:consistency}, under which the chosen state attains the maximum impact score), not consistency in the sense of a joint probability distribution over scenarios satisfying elicited constraints, as in those methods. When representativeness or scenario-set consistency is discussed in the broader literature (e.g.\ \citet{ERIKSSON2022102939}), that may refer to transparency, plausibility, or persistence of scenarios. Here the term is used only in the CIB sense. State shares or proportions reported in the results (e.g.\ in timeline figures) are empirical frequencies over pathway runs, not scenario probabilities from a single joint distribution. The present approach does not provide scenario probabilities or conservative risk bounds in the sense of \citet{Salo2022}.} Equality or near-equality within a small numerical tolerance is treated as a tie (consistent). That is, for each descriptor $j$ and for every state $l$ of that descriptor,

\begin{equation}
\label{eq:consistency}
\theta_{j,z_j}(z) \ge \theta_{j,l}(z) \quad \text{for all } l.
\end{equation}

\noindent The succession operator produces from a scenario $z$ a successor scenario $z'$ by setting each descriptor to a state that attains the maximum impact score given $z$. That is, for each descriptor $j$, $z'_j \in \argmax_l \theta_{j,l}(z)$ (with a tie-breaking rule when the maximum is not unique). \\

\noindent Starting from an initial scenario, succession is applied repeatedly. If after some number of steps the scenario equals its successor, a fixed-point attractor has been reached. If the successor was already encountered at an earlier step, a cycle attractor is identified, namely a finite repeating sequence of scenarios. Formally, a fixed point satisfies $z' = z$. A cycle is a list of scenarios $(z^{(1)}, \ldots, z^{(m)})$ such that the successor of $z^{(m)}$ is $z^{(1)}$.\\

\noindent The implementation may cap within-period iteration at a maximum of $K_{\max}$ steps. When that cap is applied and neither a fixed point nor a cycle has been reached within $K_{\max}$ steps, the last scenario $z^{(K_{\max})}$ is taken as the period outcome (partial convergence). That scenario need not satisfy Eq.~\eqref{eq:consistency}. In this analysis the succession operator is local, in that one descriptor is updated per step (the one with the largest gap between current and maximum impact score), so each step counts as a single-descriptor update.\\

\noindent In dynamic runs \footnote{Dynamic CIB is not dynamics in the sense economists validate against data. Used honestly, it is structured expert judgement stepped through time on a thin cross-impact scaffold for exploration and stress-testing. It is not a state-space model with identified propagation, micro-founded adjustment, or projections that assume a validated mapping from states to observations. Mistaking elicitation for identification, or treating these trajectories as policy-grade outputs, overstates what the scaffold can bear. Parts of the CIB literature have also pushed the method toward structural policy appraisal and similar uses where the elicited layer cannot carry the burden. Where the same artefacts are presented as quantitative dynamic analysis of real adjustment, that use is inadmissible by ordinary standards for structural dynamics and policy appraisal.} , a time grid $T$ of periods is defined (e.g.\ $T = \{t_1, \ldots, t_N\}$). An element $t \in T$ is referred to as a period \citep{ROSSROSS}. A pathway is the sequence of scenarios $(z(t))_{t \in T}$ realised at each period. Within each period, the model applies succession from the current state (or from an initial scenario in the first period), optionally after cyclic transitions for selected descriptors and threshold-based modifications to the active matrix.\footnote{Threshold-triggered CIM changes and cyclic descriptors (state in the current period dependent on the previous period) are implemented in PyCIB in line with the non-linear transformation pathway approach of e.g. \citet{vogele2019}. In that approach, a tree of consistent scenarios is built per sub-period. Here an ensemble of single-path simulations is run, with disequilibrium, regimes, and uncertain CIM added.} Structural and dynamic shocks (random perturbations to the CIM and to impact scores during succession, respectively) are described in a later subsection and are used alongside these extensions in this analysis.\\

\subsection{Disequilibrium}
\label{sec2:disequilibrium}

\noindent This extension formalises the quantification of short-lived departures from equilibrium along pathways (Section~\ref{sec:conceptual}). The following metrics define the implementation. Disequilibrium is defined here as the situation in which the realised scenario at a period is not fully characterised by a single equilibrium condition under the active matrix. That is, the state actually produced may be transiently inconsistent, in that at least one descriptor is not in the state that maximises impact balance for that period's matrix. For a scenario $z$ and descriptor $j$, the consistency margin (the margin by which the chosen state leads or lags the best alternative) is

\begin{equation}
\label{eq:margin_descriptor}
\mu_j(z) = \theta_{j,z_j}(z) - \max_{l \neq z_j} \theta_{j,l}(z),
\end{equation}

where $\theta_{j,l}(z)$ is the impact score defined in Eq.~\eqref{eq:impact}. The global consistency margin is

\begin{equation}
\label{eq:margin_global}
\mu(z) = \min_j \mu_j(z);
\end{equation}

\noindent If descriptor $j$ has only one state, the set $\{l \neq z_j\}$ is empty. In that case $\mu_j(z)$ is taken as zero (and such descriptors are included in the minimum in Eq.~\eqref{eq:margin_global}). The scenario is consistent under the active matrix if and only if $\mu(z) \ge 0$. When within-period succession is capped (partial convergence), the period outcome may have $\mu(z) < 0$.\\

\noindent The implementation quantifies disequilibrium further using distances and a time-to-consistent measure. The Hamming distance between two scenarios $z$ and $z'$ (same descriptor set) is the number of descriptors in which they differ (hence $d_{\mathrm{H}}$ is integer-valued, and the implementation may report it as a float).

\begin{equation}
\label{eq:hamming}
d_{\mathrm{H}}(z, z') = \sum_j \mathbf{1}[z_j \neq z'_j].
\end{equation}

\noindent For a scenario $z$ and the active matrix, the distance to the nearest attractor is the minimum of $d_{\mathrm{H}}(z, z^*)$ over all scenarios $z^*$ that are fixed-point attractors or belong to a cycle attractor. When exact enumeration of attractors is infeasible, the implementation reports the Hamming distance from $z$ to the attractor reached by succession from $z$ (rather than the minimum over all attractors). The distance to the consistent set is the minimum of $d_{\mathrm{H}}(z, z^{\mathrm{c}})$ over all scenarios $z^{\mathrm{c}}$ that satisfy Eq.~\eqref{eq:consistency}. When exact enumeration of consistent scenarios is infeasible (e.g.\ in this application, $3^{15}$ scenarios), the implementation reports the Hamming distance from $z$ to the first consistent scenario reached by succession from $z$.\\

\noindent The time to the consistent set is the smallest number of succession steps $k \ge 0$ such that the scenario reached after $k$ steps is consistent. If no such $k$ exists within the iteration cap, the quantity is undefined. The implementation reports this quantity as ``time to equilibrium'' in the pathway diagnostics. The same diagnostics may label the Hamming distance to the attractor reached by succession as ``distance to equilibrium'' (distinct from the distance to the consistent set). The cumulative disequilibrium burden along a pathway is

\begin{equation}
\label{eq:burden}
B\bigl((z(t))_{t \in T}\bigr) = \sum_{t \in T} \max\bigl(0, -\mu(z(t))\bigr).
\end{equation}

\noindent These metrics (margins, distances, time to consistent set, burden) are computed per period from the realised scenario and the active matrix and reported alongside the realised pathway.\\

\subsection{Extremes}
\label{sec2:extremes}

\noindent As outlined in Section~\ref{sec:conceptual}, pathways can be run under different assumed regimes. The extension is implemented as regime-aware dynamics. The model admits multiple cross-impact matrices in this setting. The user defines a set of named regimes, each associated with its own CIM, so there are as many distinct CIMs as regimes.\\

\noindent A regime is a named structural mode. Each regime $r$ has a base matrix $M_{\mathrm{base}}^{(r)}$ (or uses a shared fallback) and an optional ordered list of modifiers. The active matrix $M_r$ for regime $r$ is obtained by applying those modifiers in sequence to the base matrix.

\begin{equation}
\label{eq:regime_matrix}
M_r = (\varphi_L \circ \cdots \circ \varphi_1)\bigl(M_{\mathrm{base}}^{(r)}\bigr), \quad L \ge 1; \qquad M_r = M_{\mathrm{base}}^{(r)} \quad \text{when } L = 0.
\end{equation}

\noindent Here $\circ$ denotes composition: $(\varphi_k \circ \varphi_{k-1})(M)$ is the matrix obtained by applying $\varphi_{k-1}$ to $M$, then $\varphi_k$ to the result, so that the modifiers are applied in order $\varphi_1$, then $\varphi_2$, \ldots, then $\varphi_L$. For each period $t \in T$, the model uses the active matrix of the current regime for impact scoring and succession, and all consistency and disequilibrium measures for that period refer to $M_{r(t)}$. In the simplest configuration, the run is assigned an initial regime $r_0$ and the regime is held fixed for the whole pathway, with $r(t) = r_0$ for all $t \in T$, so the pathway answers the question of what happens when the world follows that regime's influence structure.\footnote{Alternatively, a regime transition rule can be supplied. At each period the rule may depend on the current scenario and path history to select $r(t)$, so the pathway may switch between regimes and hence between CIMs over time. The implementation records the active regime and matrix provenance per period so that results can be attributed to the chosen regime.}\\

\subsection{Unexpected outcomes}
\label{sec2:unexpected}

\noindent The influence table can be treated as uncertain (Section~\ref{sec:conceptual}), so that many plausible matrices are tried across runs and the full range of pathway endpoints, including rare or surprising ones, is visible. Unexpected outcomes are pathway endpoints or trajectories that arise when the matrix is sampled from an uncertain specification \citep{Walker01032003,KWAKKEL2013419}. \\

\noindent Structural uncertainty is implemented by sampling the CIM and running pathways, in contrast to approaches that resolve expert inconsistency via optimisation to a single consistent distribution \citep{Salo2022,Roponen2024}.\footnote{In those approaches, a ``consistent distribution'' is a single joint probability distribution over all scenarios that satisfies the elicited constraints. Here, uncertainty is represented by sampling many CIMs and running pathways, with no single joint over scenarios.} When the matrix supports sampling (e.g.\ from judgement distributions with an explicit uncertainty or confidence encoding), each matrix cell is drawn independently from a distribution around a point estimate. \\

\noindent A cell is indexed by the source descriptor, source state, target descriptor, and target state. The point estimate for that cell is $\bar{c}$, and the scale parameter for its confidence code $c$ (e.g.\ derived from expert confidence or from a prior) is $\sigma(c)$. An optional multiplicative scale $\lambda$, which may depend on run or period (e.g.\ $\lambda(t)$), scales the spread of that distribution. In the implementation used here, the period schedule is linear in the period index, $\lambda(t_i)=1+i$ for $i=0,\ldots,N-1$. The sampled value for the cell is then

\begin{equation}
\label{eq:sampling}
\tilde{c} = \mathrm{clip}\bigl( \mathcal{N}(\bar{c}, (\sigma(c) \cdot \lambda)^2), \, c_{\min}, \, c_{\max} \bigr),
\end{equation}

where $\mathcal{N}(\bar{c}, (\sigma(c)\cdot\lambda)^2)$ denotes the normal distribution with mean $\bar{c}$ and variance $(\sigma(c)\cdot\lambda)^2$ (standard deviation $\sigma(c)\cdot\lambda$), and $\mathrm{clip}(\cdot, c_{\min}, c_{\max})$ denotes clipping to the admissible CIM score range $[c_{\min}, c_{\max}]$ (when the scale is unbounded, clipping may be omitted). The full sampled matrix $\tilde{C}$ is built by drawing every cell in this way (with a specified random seed or stream for reproducibility). Different runs therefore realise different influence structures. Combined with succession and optional regime or shock mechanisms, this produces an ensemble of pathways whose distribution may include states that would be rare under the central matrix. Here the source of tail risk is uncertainty in the influence structure (Section~\ref{sec:conceptual}); the ensemble can expose both grey- and black-swan type outcomes \citep{Taleb2007}.\\

\noindent Reporting is at pathway level. The analyst inspects the distribution of endpoint states or of full trajectories across runs to quantify the share of runs that realise each outcome. Structural and dynamic shocks (additive perturbations to the matrix and to impact scores during succession) can be combined with sampled matrices for stress-testing.\\

\subsection{Exogenous stochastic shocks}
\label{sec2:shocks_and_extensions}

\noindent This analysis uses two kinds of exogenous stochastic shocks that are implemented and outlined in \citet{ROSSROSS}. Structural shocks are random perturbations applied to the cross-impact matrix. The influence table is additively (or otherwise) disturbed, then used for scoring and succession in that period or run. Structural shocks thus represent unexpected changes to the system's influence structure. Dynamic shocks, by contrast, perturb the balance for each descriptor when computing the successor scenario (e.g.\ by a draw from an autoregressive process over periods), so that the step-by-step dynamics are disturbed rather than the matrix itself. \\

\noindent Te perturbation at each descriptor-state and period follows a first-order autoregressive (AR(1)) process: $\eta_t = \rho \eta_{t-1} + u_t$ with $u_t \sim \mathcal{N}(0, (1-\rho^2)\tau^2)$, so that the long-run standard deviation is $\tau$. Both shock types are exogenous and stochastic and serve to stress-test pathways under random disturbances. In this analysis, structural and dynamic shocks are used where appropriate alongside the three extensions described in the preceding subsections, and the extensions add complementary diagnostics and scenario dimensions that the shocks do not themselves provide.\\

\section{Simulation scenarios}
\label{sec2:simulation}

\noindent The following describes the specification used in the empirical application. The general method above is independent of these choices. The approach would work with any cross-impact matrix. The illustrative example here focuses on socio-technical decarbonisation pathways and uses data sourced from \cite{ROSSROSS}. Impact scores are defined on a bounded scale and, in this application, that scale is the interval $[-3, +3]$ (as per the underlying data). Any sampled or perturbed cell value is clipped to this range. The uncertain-CIM variant uses confidence-coded expert judgements \citep{ROSSROSS}. Each cell has a point estimate $\bar{c}$ and a confidence code $c$ that maps to a standard-deviation scale $\sigma(c)$, so that the sampled value for the cell is drawn from $\mathcal{N}(\bar{c}, (\sigma(c) \cdot \lambda)^2)$ and then clipped to $[-3, +3]$.\\

\noindent The simulation runs four workflows. The full set of parameter and workflow choices, including the data source, time grid, initial scenario, run count, partial convergence and succession settings, structural and dynamic shock parameters, regime definitions, and the unexpected-outcome threshold, together with notation and assumed values, are given in \ref{appA} and \ref{appB}. Table~\ref{tab:appB_notation} defines the notation used in the main text and in \ref{appA} and, where applicable, gives the simulation assumed values.\\

\section{Results}
\label{Simulation_results}

\begin{figure}[!ht]
    \centerline{%
    \includegraphics[keepaspectratio]{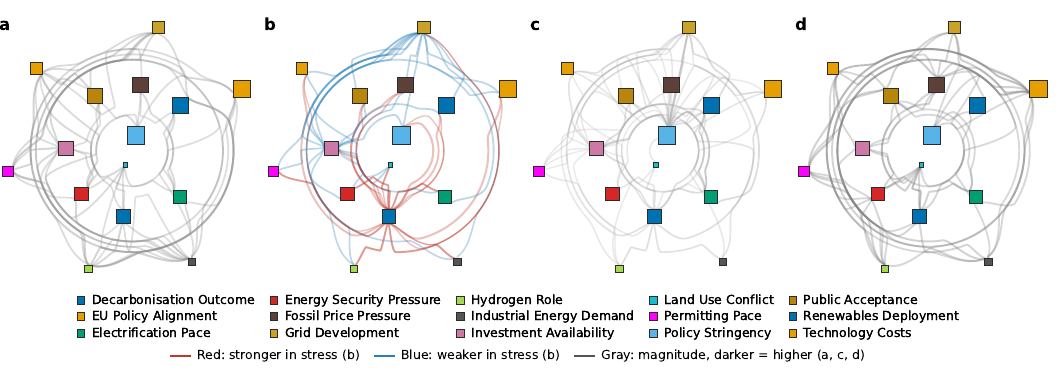}
    }
    \caption{Structural view of the cross-impact matrix under four contrasts (same layout and node-size scale, out-strength from the base CIM; edges in the upper quartile by weight, i.e.\ at or above the 75th percentile, are shown). \textbf{a}, Structural-shock illustration: mean absolute confidence-calibrated structural shock per directed edge (over state pairs), one draw; edge width and darkness show magnitude. \textbf{b}, Extremes: absolute impact difference between stress and base regimes, $|\text{stress}-\text{base}|$. \textbf{c}, Regime spread: range of mean-absolute impact across base, moderate, and high CIMs. \textbf{d}, Unexpected: standard deviation of impact strength across samples from the uncertain matrix (same edge width and darkness convention as \textbf{a}).}
    \label{fig:FIG01}
\end{figure}

\begin{figure}[!ht]
    \centerline{%
    \includegraphics[keepaspectratio]{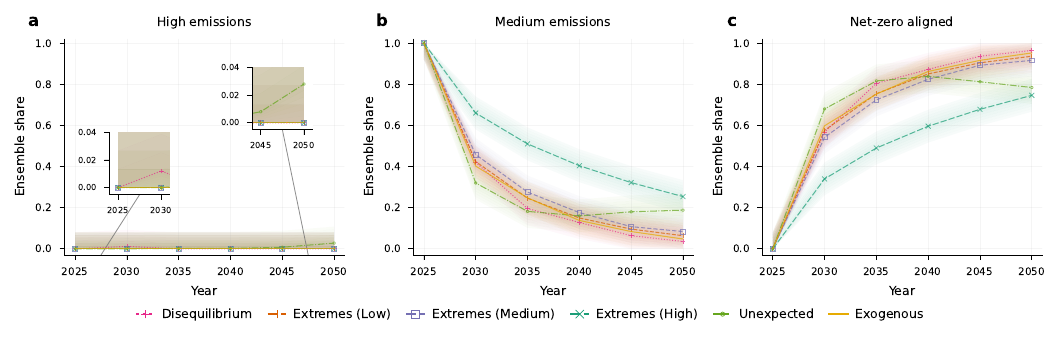}
    }
    \caption{Pathway ensemble shares for the timeline descriptor (Decarbonisation Outcome): one state per panel (a-c), all scenario types overlaid. \textbf{a}, High emissions; \textbf{b}, Medium emissions; \textbf{c}, Net-zero aligned. Lines: Disequilibrium (nominal backbone CIM, structural shocks, amplified event period); Extremes low, medium, high (nominal regime-specific CIM paired with regime-specific scaling $\alpha$, no switching); Unexpected (uncertain CIM, per-period re-sampling with $\lambda(t)$, dynamic and structural shocks); Exogenous (nominal backbone CIM, structural and dynamic shocks; no regime or sampling uncertainty). Lines are ensemble proportions; shaded bands are derived from 99\% Wilson score intervals at each period ($n = 500$ runs per type), drawn as nested layers.}
    \label{fig:FIG02}
\end{figure}

\noindent Outcomes from four scenario types are reported in this section, each corresponding to a distinct combination of assumptions and disturbances. A selection of the results is presented in two figures. Figure~\ref{fig:FIG01} provides a structural view of the cross-impact matrix. All descriptors and their pairwise influences are shown in a single layout, with four panels that differ by scenario type. To recall, the descriptors are not standalone, in that they interact through the cross-impact matrix, so that the state of each descriptor influences the impact balance of others, and succession updates the scenario step by step. The system is therefore changing over time along each pathway. Figure~\ref{fig:FIG02} focuses on a single descriptor, the timeline descriptor (Decarbonisation Outcome), and shows how the ensemble share of each of its three states evolves over time across pathway runs. In each panel, one line is plotted per run family (with three lines for the extremes family: low, medium, high), so that pathway outcomes under the different assumptions can be compared across the time horizon.\\

\noindent Outcomes differ between scenario types because each combines a different set of mechanisms. These concern whether the influence structure is sampled as uncertain, held to a nominal regime-specific backbone, or held to a single nominal backbone subject to structural and dynamic shocks, together with which shock types and amplification rules are applied. Each of the four scenario types can be put in the form of a simple question (e.g.\ ``What if one year is stressed much more strongly than the others?'' versus ``What if alternative joint regime-and-shock specifications are contrasted, for example a smoother transition against stronger headwinds?'' versus ``What if the strength with which factors affect one another is only partly known?'' versus ``What if a single nominal influence picture is assumed, whilst structural and dynamic shocks keep perturbing pathways?''). In order, these questions map to the scenario types named in the legend of Figure~\ref{fig:FIG02}, namely Disequilibrium, Extremes (low, medium, and high; regime keys), Unexpected, and Exogenous. This yields outcome distributions that are interpretable. That is, differences between scenario types can be qualitatively attributed to the combination of mechanisms in play, and both the extent of outcome variation and the scenario logic that produced it are visible.\\

\noindent Figure~\ref{fig:FIG01} shows where variation arises in the influence structure. Node size in each panel encodes out-strength from the base CIM (the sum of absolute impact strengths on outgoing edges from each descriptor). Panel (a) highlights edges most affected by a calibrated structural-shock draw (an illustration aligned with, but not parameter-identical to, the disequilibrium workflow). Panel (b) shows where the stress regime differs from the base. Panel (c) shows where regime spread is largest across the three regime specifications (base, moderate, stress). Panel (d) shows where pathway outcomes are most sensitive to CIM sampling under uncertainty. \\

\noindent Figure~\ref{fig:FIG02} compares pathway outcomes across scenario types.\footnote{Do not mix up two uses of ``medium''. The panel headings state which decarbonisation outcome is counted (High, Medium, or Net-zero). The legend states which batch of simulations each line comes from. The word ``medium'' in the extremes legend names one batch (the moderate-storyline matrix), not the Medium panel on its own.} The three panels give the ensemble share of each state of the timeline descriptor (Decarbonisation Outcome) over time for each scenario type, with 99\% Wilson score intervals (confidence intervals for binomial proportions that remain well-behaved near zero or one) over 500 runs. Stress-testing is reflected in the disequilibrium and exogenous panels. The former includes one amplified event period (2030), so transient response to a stronger perturbation is visible. The latter highlights structural and dynamic shocks about a nominal backbone CIM without uncertain-CIM sampling or regime change.\\

\noindent The sharp 2030 increase in Net-zero aligned ensemble share visible across all scenario types reflects a common baseline dynamic. All workflows use the same initial condition (middle ordinal state per descriptor), the same period grid and succession settings, and are anchored in a baseline CIM whose attractor structure is pro-transition (\ref{appA}). It follows that early-period movement is broadly aligned across workflows, whereas the scenario-specific mechanisms (event amplification, regime-specific CIM and $\alpha$, uncertain-CIM sampling, and dynamic shocks) primarily determine the extent of spread, persistence, and tail outcomes thereafter. The same shared pull also helps explain why High emissions remains near zero in most workflows across the horizon. Under the common initialisation and succession setup, trajectories are directed toward the dominant pro-transition basin; in this calibration, non-zero High-emissions mass is concentrated in the unexpected-outcomes workflow as a small late tail, while the other workflows remain at or near zero.\\

\noindent The three extremes ensembles (low, medium, and high) each combine a regime-specific CIM with regime-specific structural scaling $\alpha$. The CIMs are baseline point estimates, a moderate scaled copy, and a stress scaled copy. It must be noted that these ensembles do not isolate the effect of narrative alone. Divergence among the lines reflects joint change in influence structure and shock intensity. When the three lines differ materially, outcomes are sensitive to that combined extremes specification. When they remain close, outcomes are robust across those three regimes. The labels baseline, faster transition, and stress or constraint describe the intended storyline of each CIM variant. They do not correspond to a design in which structural scaling $\alpha$ is held fixed whilst only the storyline matrix changes.\\ 

\noindent The unexpected-outcomes scenario (uncertain CIM, re-sampled per period with period-widening $\lambda(t)$, plus dynamic and structural shocks) yields a distribution of final states across the ensemble. In the final simulation run, 2.8\% of pathways end in High emissions, which is the only final state classified as unexpected in this ensemble (empirical final-state share below 0.15 and strictly positive). Recall, here ``unexpected'' means rare in this Monte Carlo ensemble under that threshold. It is not an externally validated tail probability under the central matrix. That share is non-negligible and supports discussion of tail risk and of endpoints that a single point-estimate matrix would underrepresent once sampling and shocks are active.\\

\noindent The three-way distinction (robust across the three extremes regimes vs.\ sensitive to that joint specification vs.\ uncertainty-driven) can thus be read from the figure and from the unexpected-outcomes classification, bearing in mind that sensitivity to the extremes illustration mixes CIM variant and structural scaling $\alpha$. Under partial convergence, the disequilibrium run exhibits a mean fraction of 0.43 of periods that are inconsistent. This indicates that pathways are in transient disequilibrium in a substantial share of period outcomes, with a mean cumulative disequilibrium burden of 16.5. The burden is the sum over periods of the negative consistency margin when inconsistent, so this value indicates cumulative stress along pathways and confirms that the within-period cap prevents full convergence in many periods.\\

\noindent It must be noted that the same initial scenario, time grid, run count, baseline structural $\sigma_{\mathrm{struct}}$, and convergence settings are held constant across workflows, while selected shock settings are workflow-specific by design, so that differences in the plotted outcome distributions reflect the scenario type (mechanism) under a controlled setup. The distinction drawn in the results between the four scenario types is qualitative. The numerical difference between two scenario types, however, is not interpretable as the quantitative effect of a single factor. Each workflow differs in several respects at once, so the difference is a compound of all of them. In particular, the three extremes lines differ in both CIM and $\alpha$ simultaneously. Furthermore, many parameter choices (e.g.\ structural alphas, event period and multiplier, dynamic shock parameters, unexpected threshold, storyline scaling factors for moderate and stress CIMs) are conventional or illustrative. Results and any comparison across scenario types are conditional on those choices and are not asserted to be robust to re-parameterisation. Full specification and brief notes on parameter sources and sensitivity are given in \ref{appA}.\\

%%%%%%%%%%%%%%%%%%%%%%%%%%%%%%%%%%%%%%%%%%%%%%%%%%%%%%%%%%%%%%%%%%%%%%%%%%%%%%%%%%%%%%%%%%%%%%%%%%%%%%%%%%%%%%%%%%%%%%%%%%%%%%%%%%%%%%%%%%%%%%%%%%
\section{Discussion and Conclusions}
\label{Discussion}

\noindent This paper has introduced three extensions that are new to CIB pathway analysis: formalising disequilibrium along pathways, comparing pathways across regimes represented by alternative cross-impact matrices (the extremes illustration uses a joint regime-and-shock setup), and surfacing rare endpoints by sampling from an uncertain matrix specification. These are organised together with exogenous shocks into four scenario types in one framework. The types line up with different notions of tail or rare outcomes in the sense discussed in the results. Used together, they make both the spread of outcomes and the storyline behind each run easier to discuss; they do not by themselves deliver a quantitative attribution of differences to a single lever. Unknown unknowns are addressed here only within the current descriptor set and influence structure, so that endpoints that a central matrix may underweight can still be examined. Exploring missing factors or relations (e.g.\ omitted descriptors or interactions) is outside the scope of this paper.\footnote{That dimension could be pursued by eliciting or discovering additional descriptors (e.g.\ via horizon scanning, stakeholder workshops, or scenario discovery on quantitative models), by testing alternative or sparse influence structures to see which relations materially change pathway outcomes, or by extending uncertain-CIM sampling so that candidate descriptors or links are included in some runs and excluded in others. Such extensions are left for future work. Nevertheless, given the seminal contribution by \citet{ROSSROSS}, the cost of generating additional expert-elicited data has decreased, making such potential extensions feasible.} \\

\noindent This combination is not available in one place in the existing scenario and CIB pathway literature: four scenario types in a single framework, pathway-level disequilibrium diagnostics, and sampling from an uncertain matrix to expose rare trajectories or endpoints, plus the three-way reading of tails used in the illustration (robust across the extremes setup, sensitive to it, or driven by structural uncertainty). The implementation can be applied and extended to other cross-impact matrices and CIB applications.\\

\noindent For analysts, the main benefit is transparency of design. Instead of one band that mixes shocks, competing storylines, and doubt about how factors interact, the types differ in what is switched on in the run, so discussion can track those distinctions. Stress-testing is explicit. Comparison across the three extremes lines and the unexpected-outcomes classification supports robustness and tail conversations without relying on a single central run. Disequilibrium and exogenous panels show response to an amplified period and to ongoing shocks. Where partial convergence applies, pathway-level diagnostics such as disequilibrium burden and the fraction of periods inconsistent summarise transient stress alongside ensemble outcomes. In the illustration, regime matrices and structural scaling move together. Separating narrative from shock intensity would mean holding one fixed while varying the other in a further design.\\

\noindent Much of the methodological literature on dynamic CIB has avoided an explicit acknowledgment of a structural tension, namely that dynamics are layered onto a representation whose natural unit is a self-consistent scenario on a small, workshop-sized factor list. In that sense, dynamic CIB can force trajectory-like behaviour onto a scaffold that was never intended to encode a complete state space or governing equations for the real system. Elicitation cost is what typically keeps that scaffold compact, and adding periods, shocks, or regimes does not remove that constraint. Outputs are therefore best read as coherence and propagation of expert-structured judgements over those coordinates. The extensions developed here mitigate some of the resulting epistemic risks and improve transparency about what is and is not being claimed about ``the'' system, while still leaving omitted descriptors and mechanisms for separate elicitation or model linkage.\\

%%%%%%%%%%%%%%%%%%%%%%%%%%%%%%%%%%%%%%%%%%%%%%%%%%%%%%%%%%%%%%%%%%%%%%%%%%%%%%%%%%%%%%%%%%%%%%%%%%%%%%%%%%%%%%%%%%%%%%%%%%%%%%%%%%%%%%%%%%%%%%%%%%
\appendix
%%%%%%%%%%%%%%%%%%%%%%%%%%%%%%%%%%%%%%%%%%%%%%%%%%%%%%%%%%%%%%%%%%%%%%%%%%%%%%%%%%%%%%%%%%%%%%%%%%%%%%%%%%%%%%%%%%%%%%%%%%%%%%%%%%%%%%%%%%%%%%%%%%
\section[Appendix A]{}
\label{appA}

\noindent This appendix specifies the parameter and workflow choices used in the simulation. Notation is as in Section~\ref{sec2} and Table~\ref{tab:appB_notation}. As in Section~\ref{sec2:simulation}, the admissible CIM score range $[c_{\min}, c_{\max}]$ (the clip interval in Eq.~\eqref{eq:sampling}) is $[-3, +3]$, and sampled or perturbed cell values are clipped to this range. The exact values and code are available from the repository cited in Section~\ref{sec2}. Where a value is derived from the CIM data (e.g.\ $\sigma_{\mathrm{struct}}$ from confidence codes) or from the CIM structure (e.g.\ $K_{\max}$, initial scenario), it is noted below; other numerical choices are conventional or illustrative. The caveat in the Results (Section~\ref{Simulation_results}) applies: results and comparisons are conditional on these choices and are not asserted to be robust to re-parameterisation.\\

\noindent Table~\ref{tab:appA_workflows} summarises the four simulation workflows at a glance (CIM sampling, regimes, structural and dynamic shocks, and partial convergence). The rows should be read alongside the narrative below and with Table~\ref{tab:appB_notation} for symbols and numerical values.\\

\begin{table}[htbp]
\centering
\caption{Simulation workflows and main options.}
\label{tab:appA_workflows}
\footnotesize
\begin{tabular}{lcccc}
\hline
Option & Disequilibrium & Extremes & Unexpected & Exogenous \\
\hline
CIM sampling & No & No & Yes$^a$ & No \\
Regimes & No & Yes$^b$ & No & No \\
Structural shocks & Yes$^c$ & Yes$^d$ & Yes$^c$ & Yes$^c$ \\
Dynamic shocks & No & No & Yes & Yes \\
Partial convergence & Yes & Yes & Yes & Yes \\
\hline
\end{tabular}\\
\footnotesize $^a$One uncertain specification per run; active CIM re-sampled per period with $\lambda(t)$; $^b$three extremes ensembles, no regime switching; each ensemble pairs a regime-specific CIM with regime-specific structural scaling; $^c$structural shocks use multiplicative-magnitude scaling with baseline $\alpha=0.3$ (with one amplified event period in the disequilibrium workflow); $^d$in extremes, structural scaling is regime-specific.
\end{table}

\noindent The illustrative application of scenarios types makes use of CIM data from \citet{ROSSROSS}; the matrix $C$ (Section~\ref{sec2:cib}, Eq.~\eqref{eq:impact}) has 15 descriptors with three ordered states per descriptor, covering policy, technology, infrastructure and demand-side dimensions, and score and confidence columns (point estimate $\bar{c}$ and confidence code $c$ mapping to scale $\sigma(c)$ in Eq.~\eqref{eq:sampling}) are used. The baseline structural shock strength $\sigma_{\mathrm{struct}}$ (the standard-deviation scale of the random perturbation applied to the CIM in structural shocks, Section~\ref{sec2:shocks_and_extensions}) is derived from the CIM confidence codes using the median of the confidence-derived standard deviations. The timeline descriptor used for the ensemble share of each of its states over time across pathway runs and for defining unexpected final states of the pathway $(z(t))_{t \in T}$ (i.e.\ the descriptor whose state at the last period is used for unexpected-outcome classification) is Decarbonisation Outcome. Each cell supplies an impact score on the scale $[-3, +3]$ and a confidence code mapping to a scale $\sigma(c)$, so that both the baseline structural shock strength and the uncertain-CIM sampling (Eq.~\eqref{eq:sampling}) reflect the experts' stated uncertainty.\\

\noindent Pathways are defined over a discrete set of periods and require a starting scenario. The time grid $T$ (Section~\ref{sec2:cib}) consists of six periods $t_1, \ldots, t_N$ at five-year intervals: 2025, 2030, 2035, 2040, 2045, 2050. The initial scenario (the starting scenario for the pathway $(z(t))_{t \in T}$, i.e.\ the scenario from which succession is applied in the first period) is defined as the middle ordinal state of each descriptor under the predefined descriptor-state ordering used in the implementation \citep{ROSSROSS}.\\

\noindent So that transient disequilibrium and path diversity can be observed along pathways, within-period succession is capped rather than run to full convergence (Section~\ref{sec2:cib}). The within-period iteration cap $K_{\max}$ (the maximum number of succession steps per period before the period outcome is taken as $z^{(K_{\max})}$) is set to $\max(1, \lfloor n_{\mathrm{desc}}/2 \rfloor)$, where $\lfloor x \rfloor$ denotes the floor of $x$ (greatest integer not exceeding $x$). When equilibrium relaxation is applied, a separate iteration limit of 1000 steps is used (relaxation towards a scenario satisfying Eq.~\eqref{eq:consistency}). The succession operator is local succession: one descriptor is updated per step within each period (the most inconsistent, i.e.\ the one with the largest gap between its current and maximum impact score).\\

\noindent Given this structure, under attractor analysis, the baseline CIM exhibits a single dominant fixed-point attractor (pro-transition: Net-zero aligned, Strong grid, High investment) with a large basin share (the fraction of scenarios in that attractor's basin of attraction, i.e.\ that converge to that attractor under succession); under full within-period convergence, most periods would therefore end at that scenario and pathway diversity would be limited. The within-period cap therefore ensures that many periods stop before reaching an attractor.\\

\noindent Structural shocks (Section~\ref{sec2:shocks_and_extensions}, random perturbations applied to the CIM) use multiplicative-magnitude scaling. The perturbation is scaled by a factor $\alpha$; under this mode the effective shock magnitude is the draw multiplied by $1 + \alpha \lvert \bar{c}\rvert/3$ for cell point estimate $\bar{c}$. The baseline structural scaling $\alpha$ (the multiplier applied to the shock magnitude, Table~\ref{tab:appB_notation}) is 0.3 for the disequilibrium, unexpected-outcomes, and exogenous-shocks workflows. One designated period $t$ (2030) is given an amplified structural $\sigma_{\mathrm{struct}}$ (the scale of the random perturbation, multiplier 4.5 on $\sigma_{\mathrm{struct}}$) so that a transient event is visible in the disequilibrium panel. For the extremes workflow, the structural scaling $\alpha$ varies by regime $r$: 0.20 (low regime), 0.50 (medium regime), and 0.60 (high regime), chosen so that the three regime panels differ clearly in pathway spread. No descriptor-specific or state-specific scaling is applied.\\

\noindent Dynamic shocks (Section~\ref{sec2:shocks_and_extensions}, random perturbations applied to the impact scores $\theta_{j,l}(z)$ during succession) are used in the unexpected-outcomes and exogenous-shocks workflows. They are applied and discussed in \citet{ROSSROSS}. In this calibration, dynamic-shock settings are workflow-specific: for unexpected outcomes, $\tau=0.36$ and $\rho=0.65$; for exogenous shocks, $\tau=0.50$ and $\rho=0.20$ (Table~\ref{tab:appB_notation}). The innovation distribution is normal.\\

\noindent In the unexpected-outcomes workflow (Section~\ref{sec2:unexpected}), one uncertain matrix (specification) is used per run. The active CIM is re-sampled each period from that specification with multiplicative scale $\lambda(t)$, using the implemented linear schedule $\lambda(t_i)=1+i$ (equivalently: base 1.0 and additive increment 1.0 per period), so that uncertainty widens over time. Each cell is drawn as in Eq.~\eqref{eq:sampling}. An outcome is classified as unexpected when the empirical share of pathways ending in that final state (of the timeline descriptor) is below 0.15 and strictly positive (final-state share threshold, Table~\ref{tab:appB_notation}; a conventional cutoff for ``rare'' in this application). A near-miss threshold $\varepsilon = 0.25$ (for reporting outcomes just above that threshold) is used where applicable.\\

\noindent For the extremes workflow (Section~\ref{sec2:extremes}), three regimes $r$ define three separate pathway ensembles (Table~\ref{tab:appB_notation}). There is no regime switching; each run stays in one regime for the whole pathway, so $r(t) = r_0$ for all $t \in T$. In this empirical illustration, the extremes runs use nominal regime-specific CIMs together with regime-specific structural shock scaling, which is consistent with common storyline-style implementations in the existing scenario literature. This configuration is illustrative rather than exhaustive. The technical framework, however, treats regime choice, structural-shock scaling, and uncertain-CIM sampling as separable optional layers, so implementations can instead hold one layer fixed whilst varying another (e.g.\ fixed scaling across regimes, fixed regime CIM across scaling levels), or allow regime switching over time. This layer-separable implementation extends common storyline-style implementations in the current literature, even though the specific empirical configuration shown here is intentionally aligned with them.\\

\noindent In the present setup, each regime's matrix is supplied as a fixed $M_{\mathrm{base}}^{(r)}$ with no further runtime composition of modifiers ($L = 0$ in Eq.~\eqref{eq:regime_matrix}). The moderate and high matrices are deterministic multiplicative scalings of the baseline cells (Table~\ref{tab:appB_notation}), not new elicitations: moderate strengthens impacts involving policy stringency, electrification pace, and renewables deployment (as source or target); stress strengthens involvement of the decarbonisation outcome descriptor, weakens impacts that target grid development and investment availability, and applies an additional down-weighting to impacts toward selected pro-transition target states so that a stress- or constraint-leaning attractor can dominate for illustration. A separate robustness evaluation in the implementation operates on the uncertain matrix specification and is not the same object as these deterministic regime CIMs; only the pathway figures in the main text use this deterministic extremes setup.\\

\noindent Parameters derived from the CIM or its structure include: $\sigma_{\mathrm{struct}}$ (median of confidence-derived scales from the CIM), the score range $[-3,+3]$ and clip interval, the descriptor set and timeline descriptor, and the initial scenario (middle ordinal state per descriptor under the predefined ordering). The within-period cap $K_{\max} = \lfloor n_{\mathrm{desc}}/2 \rfloor$ is a design choice to allow transient disequilibrium; larger $K_{\max}$ would allow more convergence and typically reduce pathway diversity. The remaining numerical choices (e.g.\ baseline $\alpha$, event period and multiplier, workflow-specific $\tau$ and $\rho$, $\lambda(t)$ growth, unexpected-outcome threshold, run count) are conventional or illustrative.\\

\noindent No full sensitivity analysis is conducted. The focus of the paper is on pathway mechanisms (disequilibrium, regime-based extremes, uncertain influence structure, and exogenous disturbances) and on illustrating how each can be implemented and read from ensemble outputs, rather than on calibrating a single ``best'' parameter set or on quantifying robustness of numerical magnitudes across the full input space.\\

%%%%%%%%%%%%%%%%%%%%%%%%%%%%%%%%%%%%%%%%%%%%%%%%%%%%%%%%%%%%%%%%%%%%%%%%%%%%%%%%%%%%%%%%%%%%%%%%%%%%%%%%%%%%%%%%%%%%%%%%%%%%%%%%%%%%%%%%%%%%%%%%%%
\section[Appendix B]{}
\label{appB}

\footnotesize
\begin{longtable}{@{}l>{\raggedright\arraybackslash}p{0.45\linewidth}>{\raggedright\arraybackslash}p{0.28\linewidth}@{}}
\caption{Notation and assumed values.}\label{tab:appB_notation}\\
\hline
Symbol & Description & Value or source \\
\hline
\endfirsthead

\hline
Symbol & Description & Value or source \\
\hline
\endhead

\hline
\endfoot

\hline
\endlastfoot

\multicolumn{3}{l}{\textit{Scenarios and CIM}} \\
$C$ & Cross-impact matrix (CIM) & \citep{ROSSROSS} \\
$\tilde{C}$ & Sampled CIM (from uncertain specification) & — \\
$z$, $z'$ & Scenario (one state per descriptor); successor scenario & — \\
$z_j$, $z_i$, $z'_j$ & State of descriptor $j$ (or $i$) in scenario $z$; state of $j$ in successor $z'$ & — \\
$(z^{(1)}, \ldots, z^{(m)})$ & Cycle of scenarios (cycle attractor); $z^{(k)}$ scenario at step $k$ & — \\
$z^{(K_{\max})}$ & Period outcome when succession is capped (partial convergence) & — \\
$z^*$, $z^{\mathrm{c}}$ & Attractor scenario; consistent scenario & — \\
$(z(t))_{t \in T}$ & Pathway (sequence of scenarios over periods) & — \\
$i$, $j$ & Descriptor indices & — \\
$l$ & State index (possible state of a descriptor) & — \\
$m$ & Cycle length (number of scenarios in a cycle) & — \\
— & Application (this study): 15 descriptors, 3 states each with predefined ordinal descriptor-state order; timeline descriptor Decarbonisation Outcome; initial scenario middle ordinal state per descriptor & \citep{ROSSROSS} \\
\multicolumn{3}{l}{\textit{Impact and consistency}} \\
$\theta_{j,l}(z)$ & Impact score for descriptor $j$ in state $l$ given scenario $z$ (Eq.~\ref{eq:impact}) & — \\
$\theta_{j,z_j}(z)$ & Impact score at the chosen state $z_j$ for descriptor $j$ & — \\
$C_{i \to j}(z_i, l)$ & Impact of source $i$ in state $z_i$ on target $j$ in state $l$ & — \\
$\mu_j(z)$, $\mu(z)$ & Per-descriptor consistency margin; global consistency margin (Eqs.~\ref{eq:margin_descriptor}, \ref{eq:margin_global}); $\mu_j(z) = 0$ when descriptor $j$ has only one state & — \\
$\argmax_l$ & Argument of the maximum over states $l$ (tie-breaking applied if not unique) & — \\
\multicolumn{3}{l}{\textit{Time and iteration}} \\
$T$, $t$ & Time grid; single period & — \\
$t_1, \ldots, t_N$; $N$ & Periods in the grid; number of periods & 2025, 2030, 2035, 2040, 2045, 2050; 6 \\
$K_{\max}$ & Within-period succession iteration cap; relaxation limit 1000 & $\max(1, \lfloor n_{\mathrm{desc}}/2 \rfloor)$ \\
$\lfloor x \rfloor$ & Floor of $x$ (greatest integer $\le x$) & — \\
$k$ & Number of succession steps (e.g.\ time to consistent set; reported as ``time to equilibrium'' in implementation) & — \\
— & Runs per workflow; seeds & 500; fixed bases and per-run offsets \\
— & Succession operator & Local succession (one descriptor per step, the most inconsistent) \\
\multicolumn{3}{l}{\textit{Disequilibrium}} \\
$d_{\mathrm{H}}(z, z')$ & Hamming distance between two scenarios (Eq.~\ref{eq:hamming}); integer (number of descriptors in which they differ) & — \\
$\mathbf{1}[\cdot]$ & Indicator function (one if condition holds, zero otherwise) & — \\
$B(\cdot)$ & Cumulative disequilibrium burden along pathway (Eq.~\ref{eq:burden}) & — \\
\multicolumn{3}{l}{\textit{Regimes}} \\
$r$, $r(t)$, $r_0$ & Regime (named structural mode); regime at period $t$; initial regime & This study: regime keys low, medium, high (no switching) \\
$M_{\mathrm{base}}^{(r)}$, $M_r$ & Base CIM for regime $r$; active matrix for regime $r$ (Eq.~\ref{eq:regime_matrix}) & — \\
$M_{r(t)}$ & Active matrix at period $t$ (depends on current regime) & — \\
$\varphi_1, \ldots, \varphi_L$ & Modifiers applied in sequence to base matrix; $L$ number of modifiers & — \\
$\circ$ & Composition: $(f \circ g)(M)$ means apply $g$ to $M$, then $f$ to the result (Eq.~\ref{eq:regime_matrix}) & — \\
\multicolumn{3}{l}{\textit{Extremes: deterministic storyline CIM scaling}} \\
— & Moderate regime: multipliers on impacts involving Policy Stringency; Electrification Pace; Renewables Deployment (as source or target; combined multiplicatively per cell) & 1.4; 1.25; 1.20 \\
— & Stress regime: multiplier on impacts involving Decarbonisation Outcome (as source or target) & 1.5 \\
— & Stress regime: multiplier on impacts with target Grid Development or Investment Availability & 0.85 \\
— & Stress regime: extra multiplier on impacts to pro-transition targets (Decarbonisation Outcome $\to$ Net-zero aligned; Grid Development $\to$ Strong; Investment Availability $\to$ High) & 0.50 \\
\multicolumn{3}{l}{\textit{Uncertain CIM and sampling}} \\
$\bar{c}$, $c$ & Point estimate for a CIM cell; confidence code (maps to scale) & — \\
$\sigma(c)$ & Scale from confidence code $c$ for that cell & — \\
$\lambda$, $\lambda(t)$ & Multiplicative scale for CIM sampling (optional; may depend on run or period) & unexpected workflow: $\lambda(t_i)=1+i$ (base 1.0; additive increment 1.0 per period) \\
$\tilde{c}$ & Sampled cell value (Eq.~\ref{eq:sampling}) & — \\
$c_{\min}$, $c_{\max}$ & Lower and upper bounds of admissible CIM score range & $-3$, $+3$ \\
$[c_{\min}, c_{\max}]$ & Admissible CIM score range (clip interval) & $[-3, +3]$ \\
$\mathcal{N}(\mu, \sigma^2)$ & Normal distribution with mean $\mu$, variance $\sigma^2$ & — \\
$\mathrm{clip}(\cdot, a, b)$ & Clipping to interval $[a, b]$ & — \\
— & Unexpected outcome threshold (final-state share) & 0.15 \\
\multicolumn{3}{l}{\textit{Simulation and shock parameters}} \\
AR(1) & First-order autoregressive process (e.g.\ $\eta_t = \rho \eta_{t-1} + u_t$) & — \\
$\alpha$ & Structural shock scaling (baseline or by regime) & Baseline 0.3; low regime 0.20, medium regime 0.50, high regime 0.60 \\
$\sigma_{\mathrm{struct}}$, $\sigma_{\mathrm{judg}}(c)$ & Structural shock strength; judgement scale for cell $c$ (in prose) & Median of confidence-derived std.\ devs.\ from CIM \citep{ROSSROSS} \\
— & Event period; $\sigma_{\mathrm{struct}}$ multiplier (one period) & 2030; 4.5 \\
$\eta_t$, $u_t$ & Perturbation to impact balance at (descriptor, state) and period $t$; innovation in AR(1) (Section~\ref{sec2:shocks_and_extensions}) & — \\
$\tau$, $\rho$ & Long-run std.\ and persistence in AR(1) $\eta_t = \rho \eta_{t-1} + u_t$ for dynamic shocks; innovation distribution Normal & Unexpected: 0.36, 0.65; Exogenous: 0.50, 0.20 \\
$\varepsilon$ & Near-miss threshold (reporting) & 0.25 \\
$n_{\mathrm{desc}}$ & Number of descriptors & 15 \\
\end{longtable}
\normalsize

%%%%%%%%%%%%%%%%%%%%%%%%%%%%%%%%%%%%%%%%%%%%%%%%%%%%%%%%%%%%%%%%%%%%%%%%%%%%%%%%%%%%%%%%%%%%%%%%%%%%%%%%%%%%%%%%%%%%%%%%%%%%%%%%%%%%%%%%%%%%%%%%%%

\section*{Data statement} \noindent Code and data supporting the simulations and figures in this paper are available at \url{https://doi.org/10.5281/zenodo.19364824}.

\section*{Acknowledgements} \noindent Special thanks go to Fiona for insightful discussions about probability concepts, and to Alan Ross for contributions to improving the methodology. This paper is funded by the Helmholtz Association under the programme ``Energy System Design''. Open Access is funded by the Deutsche Forschungsgemeinschaft (DFG, German Research Foundation) (491111487).\\

\renewcommand{\bibsection}{%
    \section*{References} 
    \addcontentsline{toc}{section}{References}
}
\bibliographystyle{elsarticle-harv} 
\bibliography{references}

\end{document}